\documentclass[pra,twocolumn,showpacs, superscriptaddress]{revtex4}

\usepackage{color}
\usepackage{colordvi}
\usepackage{amssymb}
\usepackage{amsmath}
\usepackage{epsf}
\usepackage{mathrsfs}
\usepackage{graphicx}
\usepackage{appendix}
\usepackage{latexsym,epsfig}

\usepackage{ulem}

\def\vecr{\textit{\textbf{r}}}
\def\site{^0}

\usepackage{ulem}

\begin{document}
	
\title{Twonniers: \\ Interaction-induced effects on Bose--Hubbard parameters}
	
\author{Mark Kremer}

\affiliation{Institut f\"ur Physik, Universit\"at Rostock, Albert-Einstein-Str. 23,18059 Rostock, Germany}
\author{Rashi Sachdeva}
\email{rashi.sachdeva@oist.jp}
\affiliation{Quantum Systems Unit, Okinawa Institute of Science and Technology Graduate University, Okinawa 904-0495, Japan}
\author{Albert Benseny}
\affiliation{Quantum Systems Unit, Okinawa Institute of Science and Technology Graduate University, Okinawa 904-0495, Japan}
\author{Thomas Busch}
\affiliation{Quantum Systems Unit, Okinawa Institute of Science and Technology Graduate University, Okinawa 904-0495, Japan}

\date{\today}

\begin{abstract}
We study the effects of the repulsive on-site interactions on the broadening of the localized Wannier functions used for calculating the parameters to describe ultracold atoms in optical lattices. For this, we replace the common single-particle Wannier functions, which do not contain any information about the interactions, by two-particle Wannier functions (``Twonniers") obtained from an exact solution which takes the interactions into account. We then use these interaction-dependent basis functions to calculate the Bose--Hubbard model parameters, showing that they are substantially different both at low and high lattice depths, from the ones calculated using single-particle Wannier functions. Our results suggest that density effects are not negligible for many parameter ranges and need to be taken into account in metrology experiments.
\end{abstract}


\maketitle

\section{Introduction}

Ultracold atoms in optical lattices have been a recent topic of significant interest, as they can be used to perform quantum simulations of fundamental models of many-body physics, which are often difficult to access using traditional condensed matter systems~\cite{Greiner:02,Yukalov:09,lewenstein_book}. The perfect periodicity of  optical lattices allows to 
mimic the crystalline environments electrons experience in solids and unprecedented control over the kinetic properties of the atoms is possible by tuning the lattice depths.
Furthermore, the interaction properties between the ultracold atoms can be changed using techniques like Feshbach resonances. This has opened up many new avenues of research, particularly in the field of condensed matter and atomic physics, and made it possible to study quantum phases and quantum phase transitions over a wide range of parameters~\cite{Greiner:02, Yukalov:09,lewenstein_book, Jordens:08}. 

Theoretically, ultracold atoms in optical lattices can be described by a Bose--Hubbard model~\cite{Fisher:89, Jaksch:98, vanOosten:01, Zwerger:03}, which stems from a mapping of the continuous system to the lattice by using site localized single-particle Wannier functions.  The static and dynamics properties of the gas are then described by two main parameters: the hopping term, which accounts for bosons tunneling between neighboring sites, and the on-site interaction term, which accounts for the repulsive energy when two particles sit at the same lattice site.
The competition between these parameters (commonly determined by calculating overlap integrals using single-particle Wannier functions) characterizes the Mott-insulator/superfluid transition~\cite{Greiner:02}.

However, while mathematically convenient, single particle Wannier functions neglect certain physical effects, such as the broadening of the localized wave functions due to repulsive on-site interactions when two or more bosons occupy the same lattice site. This can have significant effects when trying to make precision measurements~\cite{Campbell:06} or when using optical lattices for metrology~\cite{latticeclock:14}, as the energy scales that govern the behavior of the atoms are typically small.

Recently, a number of theoretical efforts have been made to incorporate the effects of interaction on the Wannier functions using mean-field and numerical approaches~\cite{Li:06,Liang:09,Schneider:09, Luhmann:12,Zhu:15}. In addition, there has been strong experimental evidence of the broadening of Wannier function at high fillings, when high-resolution spectroscopy showed non-uniform frequency shifts for different occupation numbers per site ~\cite{Campbell:06}.  It is therefore important to include the effects of modified densities due to the repulsive interactions when calculating the Bose--Hubbard parameters. In this work we suggest to do this by using the exact two-particle wave functions (``Twonniers"), obtained after solving the two-particle Schr\"odinger equation with contact interaction. For comparison, we also perform calculations using the single particle Wannier functions. To the best of our knowledge, this is the first study where the expansion is directly performed in terms of the two-particle wave functions, which has an implicit dependence on repulsive atom-atom interactions.

Our presentation is organized as follows. In Sec.~\ref{sec:model} we provide a brief review of the conventional way of calculating the Hubbard parameters using the single-particle Wannier function approach.
Then, in Sec.~\ref{sec:2pwf} we introduce the two-particle wave functions that include the interaction effects by solving the two-particle Schr\"odinger equation with contact interaction.
These wave functions are used in Sec.~\ref{sec:newBH} to calculate the parameters of the modified Bose--Hubbard Hamiltonian,
which are interpreted in Sec.~\ref{sec:results} in comparison to those obtained from single-particle Wannier functions.
Finally, we discuss possible applications and conclude in Sec.~\ref{sec:conc}. 

\section{The Bose--Hubbard model} 
\label{sec:model}

The starting point for our analysis is the Hamiltonian for a Bose gas, given by 
\begin{equation}
	\label{hamsecquan}
	\hat{H} = \hat{H}_\text{SP} + \hat{H}_\text{I} ,
\end{equation} 
where the single-particle term includes the kinetic energy and the optical lattice potential,
\begin{equation}
	\label{hamsecquan_non}
	\hat{H}_\text{SP} =	\int \mathrm{d}\vecr\,\hat{\Psi}^\dagger (\vecr) \left[-\frac{\hbar^2}{2m}\nabla^2+V_\text{L}(\vecr)\right]\hat{\Psi} (\vecr).
\end{equation}
Here $m$ is the atomic mass.
The term including the point-like interactions is given by
\begin{equation}
	\label{hamsecquan_int}
	\hat{H}_\text{I} = \frac{g}{2}\int\mathrm{d}\vecr\,\hat{\Psi}^\dagger (\vecr)\hat{\Psi}^\dagger (\vecr)\hat{\Psi} (\vecr)\hat{\Psi} (\vecr) ,
\end{equation}
where $g=4\pi\hbar^2 a_s/m$ is the interaction strength related to the $s$-wave scattering length, $a_s$.
The bosonic field operators, $\hat\Psi$ and $\hat\Psi^\dagger$, can be expanded into a series of orthonormal functions, $f_i(\vecr)$, and bosonic  annihilation and creation operators, $\hat{a}_i$ and $\hat{a}_i^\dagger$, for each lattice site as
\begin{equation}
	\hat{\Psi} (\vecr)=\sum_i f_i(\vecr)\hat{a}_i
	\quad \text{with}\quad 
	\int \mathrm{d}\vecr\, 	f_i^*(\vecr)f_j(\vecr)=\delta_{ij}. \label{ortho}
\end{equation}
A convenient and common choice for the orthonormal functions in a lattice potential are the well-known Wannier functions~\cite{wan37, kohn73}, which are localized at the individual lattice sites. 
The single-particle Wannier function at lattice site $i$ in the Bloch band $\alpha$ is defined as 
\begin{equation}
	w^{\alpha}_i(\vecr) = w_{i,x}^{\alpha}(x) w_{i,y}^{\alpha}(y) w_{i,z}^{\alpha}(z) ,
	\label{wannier}
\end{equation}
and the components in each direction can be written in terms of the Bloch functions $\phi^\alpha_k(x)$ as 
\begin{equation}
	w_{i,x}^{\alpha}(x) = \frac{1}{\sqrt{N_x}}\sum_k\mathrm{e}^{-\mathrm{i} k x\site_i}\phi^\alpha_k(x), \label{wan1d}
\end{equation}
where $N_x$ is the number of lattice sites along the $x$-direction (equivalent expressions exist for the other spatial directions), and $x\site_i$ is the center of the $i$-th trap.
It is important to note that the Wannier functions are not eigenfunctions of the system and that, as single-particle functions, they do not contain any information about possible scattering effects due to multi-particle occupancy of a site.
Also, for small interaction energies the particles can be considered to be confined in the lowest Wannier orbitals because the energy separation between the lowest and first excited band is quite large compared to interaction energy. We work in this regime and from now onwards will drop the band index $\alpha$.
 
The hopping amplitude in the Bose--Hubbard model can then be calculated as
\begin{equation}
	\label{hopterm}
	J=\int \mathrm{d}\vecr\, w_i^*(\vecr) \left[-\frac{\hbar^2}{2m}\nabla^2+V(\vecr)\right]w_i(\vecr),
\end{equation}
where only the nearest-neighbor overlaps are taken into account, and the interaction part of the Hamiltonian leads to the onsite interaction amplitude
\begin{equation}
\label{onsiteterm}
U=g\int \mathrm{d}\vecr\, |w_i(\vecr)|^{4} .
\end{equation}

\section{Two-particle wave functions}
\label{sec:2pwf}

The effect of the repulsive scattering interaction depends on both the interaction strength $g$ and the density distribution of the wave function (see Eq.~\eqref{hamsecquan_int}).
Therefore, it is important to choose the correct form for the orthonormal functions with which one performs the expansion: since the interactions are local and the functions are localized the density distribution should take the  interaction into account if two (or more) particles are at the same lattice site. We will therefore in the following replace terms of the form $f_i(\vecr)f_i(\vecr)$ by two-particle Wannier functions, but leave terms of the form $f_i(\vecr)f_j(\vecr)$ $(i\neq j)$ to be described by single-particle Wannier functions. 

To find the two-particle Wannier functions we solve the Schr\"odinger equation for two particles in a sinusoidal potential, $V_\text{L}(\vecr)$, interacting via a point-like potential. The Hamiltonian is given by
\begin{equation}
	\hat{H} = \sum_{k=1}^2\left[-\frac{\hbar^2}{2m}\nabla^2_k+V_\text{L}(\vecr_k)\right]+\frac{g}{2}\delta(\vecr_1-\vecr_2),
\label{lattice}
\end{equation}
and its corresponding delocalized eigenfunctions $\Phi_j(\vecr_1,\vecr_2)$  can be used as a basis to construct the localized  (two-particle) functions
\begin{equation}
W_i(\vecr_1,\vecr_2)=\sum_j c_j\Phi_j(\vecr_1,\vecr_2) \quad \text{with} \ \sum_j|c_j|^2=1.
\end{equation}
Since the interactions raise the energies, we use the eigenfunctions of the two lowest bands.
To determine the coefficients $c_j$, we assume that the particles are well localized at each lattice site, using as the criteria for localization the minimization of the second moment~\cite{Stollenwerk:11}
\begin{equation}
	M_i=\int \mathrm{d}\vecr_1\mathrm{d}\vecr_2\,W_i^*(\vecr_1,\vecr_2)\left(\vecr_1^2+\vecr_2^2\right)W_i(\vecr_1,\vecr_2). \label{secmoment}
\end{equation}
This allows us to define the single-particle single-site densities from the two-particle wave functions as
$|W_i(\vecr,\vecr)|$. In order to fulfill the orthogonality condition in Eq.~\eqref{ortho} this density needs to be normalized as
\begin{equation}
\int \mathrm{d}\vecr\,|W_i(\vecr,\vecr)|\overset{!}{=}1,
\end{equation}
which also assures the fulfilment of the particle statistics,
\begin{equation}
	\left[a_i,a_j^\dagger\right] = \delta_{i,j}\quad\text{and}\quad\left[a_i,a_j\right]=\left[a_i^\dagger,a_j^\dagger\right]=0. \label{stats}
\end{equation}

\begin{figure}[tb]
\includegraphics[width=0.45\textwidth]{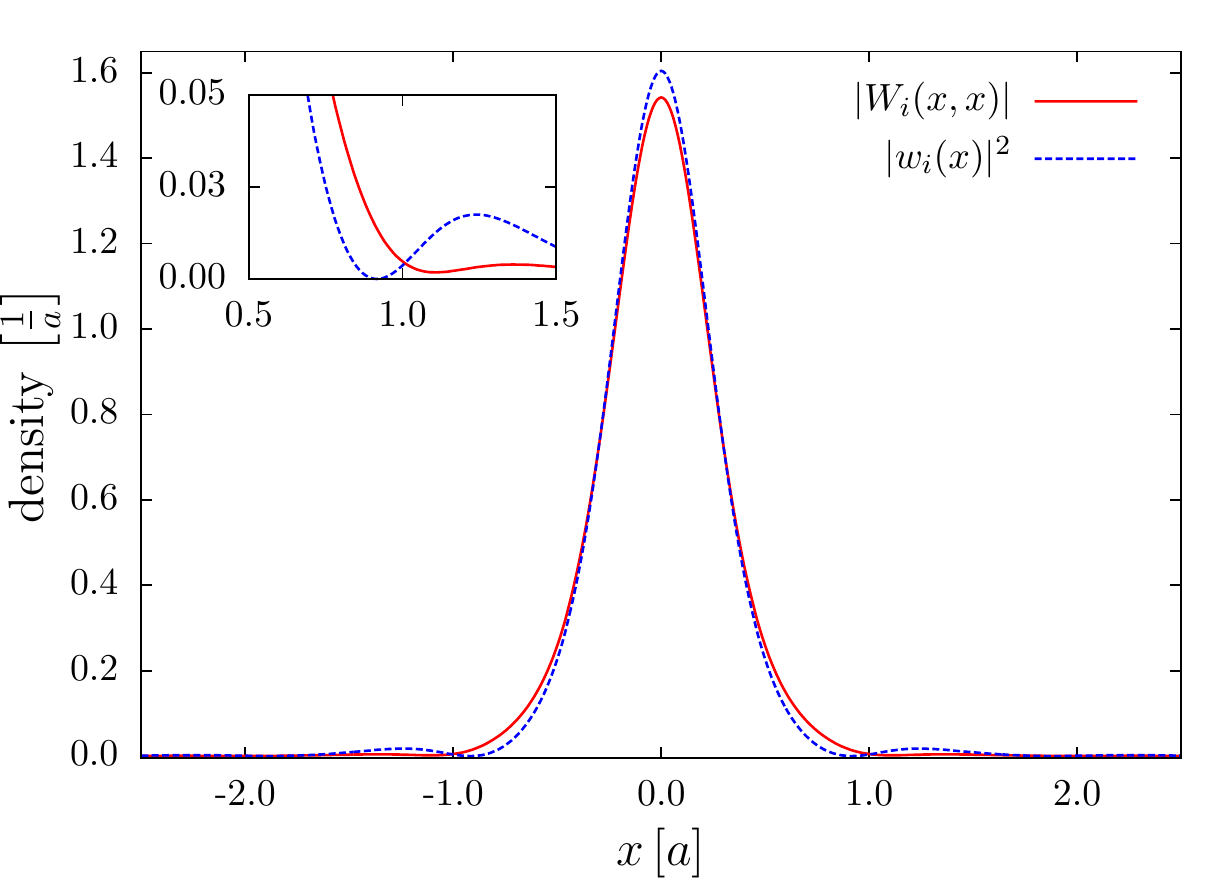}
\caption{\label{fig:comparison} The red (solid) line corresponds to the two-particle single site density obtained after numerically solving the Schr\"odinger equation with the Hamiltonian \eqref{lattice} for 9 traps, with lattice depth $V_0=1.5 E_r$ and scattering length $a_s=100a_0$, where $a_0$ is the Bohr radius.  The blue (dashed) line corresponds to the square of single-particle Wannier functions for the same lattice parameters. The lattice depth is given in units of the recoil energy $E_r=\pi^2\hbar^2/2ma^2$, where $a$ is the lattice spacing of the sinusoidal optical lattice potential.
The inset shows a zoom-in on the tails of the densities, clearly showing the broadening of two-particle density compared to the density of the single-particle Wannier function.}
\end{figure}

To compare the single particle and two-particle Wannier functions, we show in Fig.~\ref{fig:comparison} their respective densities computed in a one-dimensional  potential $V_\text{L}(x)=V_0\sin^2(\pi x/a)$. 
One can clearly see that, as expected, the repulsive interaction leads to a broadening of two-particle Wannier function, which eventually results in significant change in the Bose--Hubbard parameters.
However, one can also see that the wings of the two particle Wannier function at the position of the neighbouring lattice sites are suppressed, which is due to the orthogonality requirement between two of the modified Wannier functions.

In the next section, we use this two-particle wave function and density to construct the different terms in the Hamiltonian and compare them to the ones using only single-particle Wannier function solutions.

\begin{figure*}[tb]
\begin{minipage}{0.49\textwidth}
\centering (a) $a_s=100a_0$ \\ \includegraphics[width=0.9\textwidth]{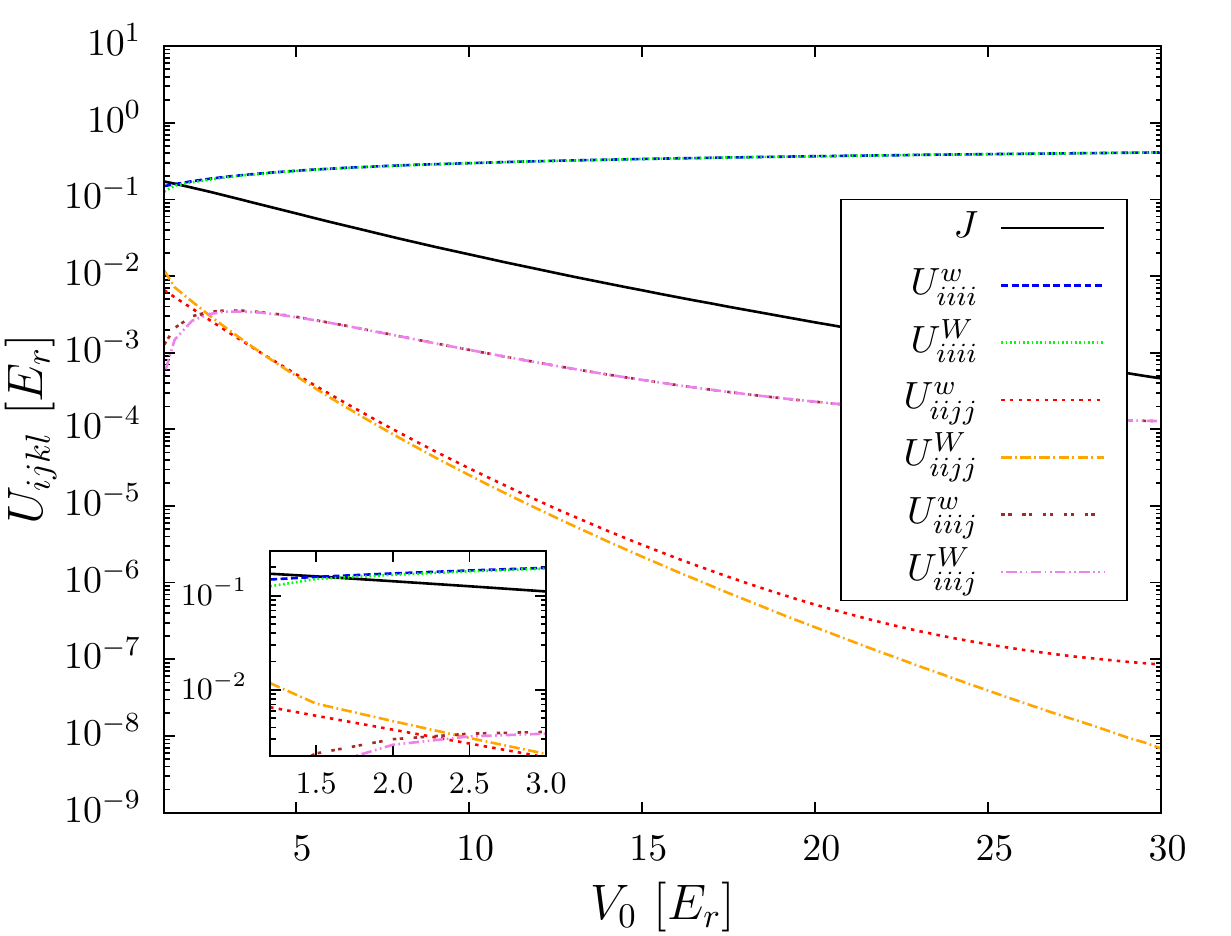}
\end{minipage}
\begin{minipage}{0.49\textwidth}
\centering (b) $a_s=400a_0$ \\ \includegraphics[width=0.9\textwidth]{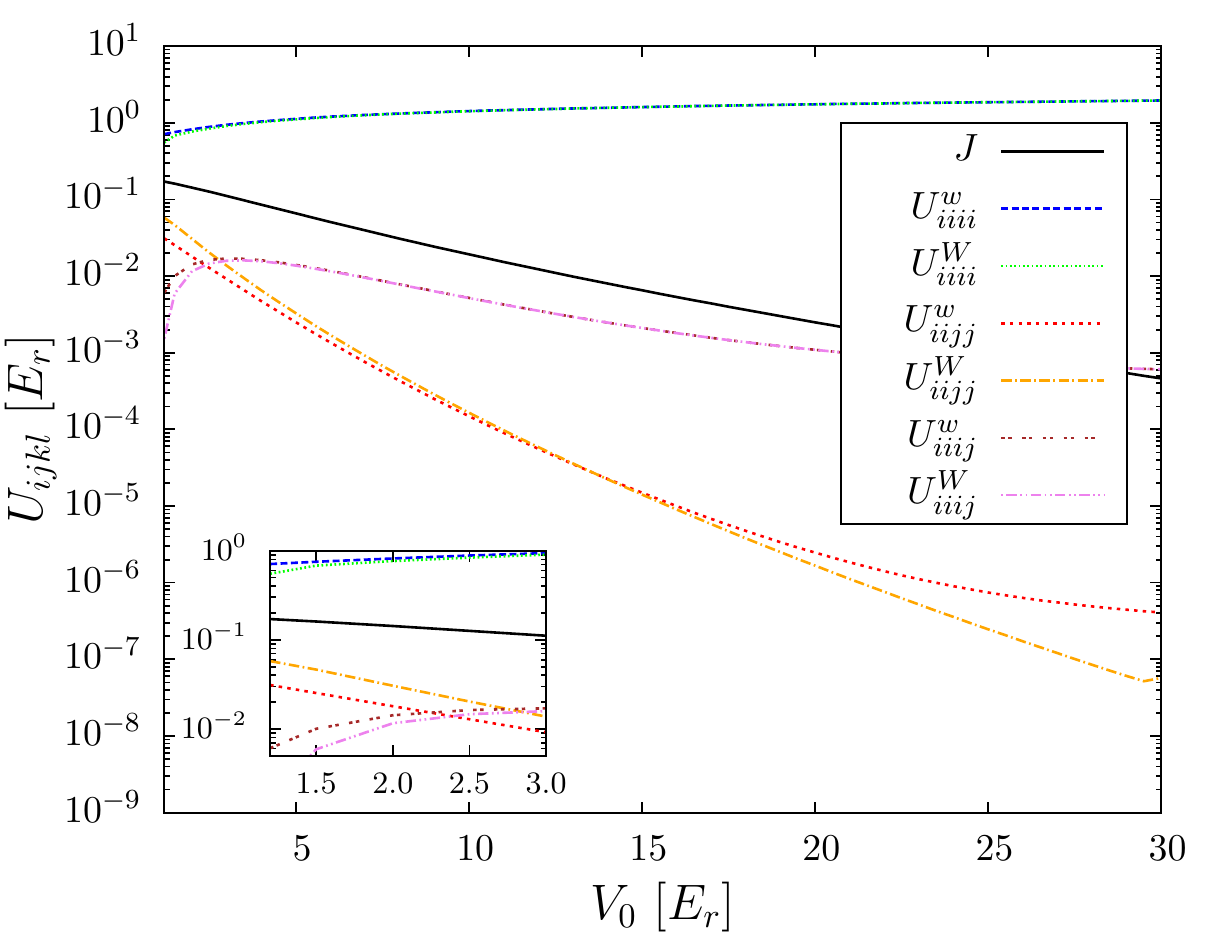}
\end{minipage}
\includegraphics[width=0.95\textwidth]{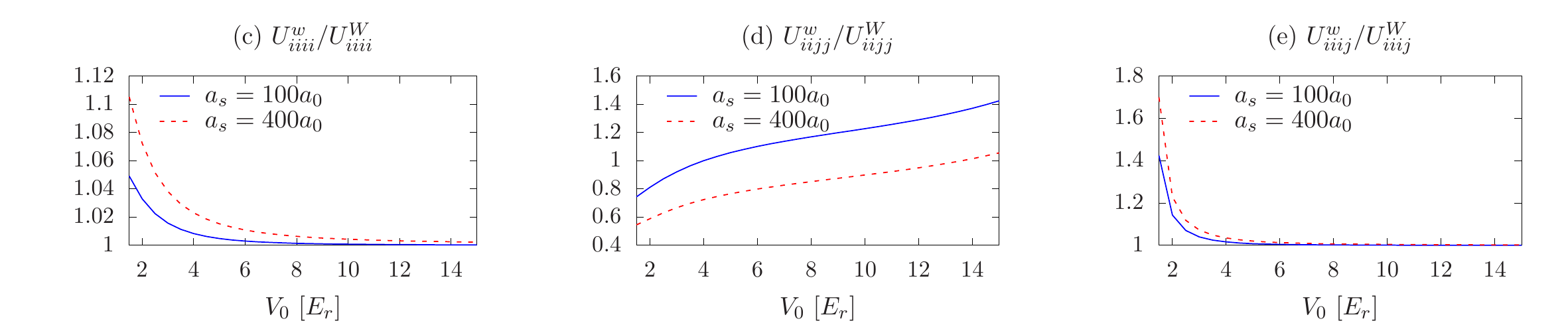}
\caption{\label{fig:depths}
Top row:
Dependence of the Bose--Hubbard parameters (plotted logarithmically) on the lattice depth $V_0$ (in units of the recoil energy $E_r=\pi^2\hbar^2/2ma^2$)
for scattering lengths (a) $a_s=100a_0$, and (b) $a_s=400a_0$, with $a_0$ being the Bohr radius.
The curves correspond to
$U_{iiii}^{w}$ (dashed blue),
$U_{iiii}^{W}$ (dotted green),
$J$ (solid black),
$U_{iiij}^{w}$ (double-dotted dark red),
$U_{iiij}^{W}$ (dot-dash-dotted pink),
$U_{iijj}^{w}$ (small-dashed red), and
$U_{iijj}^{W}$ (dashed-dotted orange). The insets shows the behavior for shallow lattices.
For the numerical calculation 9 traps have been taken into account. Bottom row:
Ratios of (c) $U_{iiii}$, (d) $U_{iijj}$, and (e) $U_{iiij}$, calculated with the two methods (single-particle and two-particle Wannier functions) for $a_s = 100 a_0$ (solid blue) and $a_s = 400 a_0$ (dashed red).
}
\end{figure*}

\section{Modified Bose--Hubbard Hamiltonian}
\label{sec:newBH}

The effects of the interactions between the particles are fully contained in the interaction term $\hat{H}_\text{I}$, which, after
inserting the expansion of Eq.~\eqref{ortho}, takes the form
\begin{align}
	\hat{H}_\text{I}&=\frac{g}{2}\sum_{ijkl}\int\mathrm{d}\vecr\, f_i^*(\vecr)f_j^*(\vecr)f_k(\vecr)f_l(\vecr)\hat{a}_i^\dagger \hat{a}_j^\dagger \hat{a}_k \hat{a}_l \nonumber\\
	&=\frac{1}{2}\sum_{ijkl}U_{ijkl}\hat{a}_i^\dagger \hat{a}_j^\dagger \hat{a}_k \hat{a}_l.
	\label{genintham}	
\end{align}
As we are only interested in the ground state, the Wannier functions and the two-particle wave functions based on Eq.~\eqref{lattice} can be chosen to be real and we will therefore neglect the complex conjugates below.  The parameters $U_{ijkl}$ can then be calculated using the substitution
\begin{equation}
\label{cond_n}
f_i(\vecr)f_j(\vecr)
\xrightarrow{\ W \ }
\begin{cases}
|W_i(\vecr,\vecr)| & \text{if $i=j$}, \\
w_i(\vecr)w_j(\vecr) & \text{if $i\neq j$},
\end{cases}
\end{equation}
which should be compared to the standard way of calculating using single-particle Wannier functions
\begin{equation}
\label{cond_w}
f_i(\vecr)f_j(\vecr)
\xrightarrow{\ w \ }
w_i(\vecr)w_j(\vecr)
\quad \forall i,j.
\end{equation}
Here we have introduced the labels $W$ and $w$ which will be used below to distinguish, respectively, terms calculated from the two-particle Wannier function density or from single-particle Wannier functions.
The hopping term in the Bose--Hubbard model depends only on the single-particle Wannier functions  as it comes from the non-interacting part of the Hamiltonian \eqref{hamsecquan_non}, and it is therefore is not affected by these substitutions. 

To explicitly identify the different physical processes that are summarized in the interaction term, we will in the following group the different terms into four categories.
The first one is the one where two particles are at the same site and interact with each other. The associated terms include $\hat{a}_i^\dagger \hat{a}_i^\dagger \hat{a}_i \hat{a}_i$ and
their corresponding amplitude is given by
\begin{equation}
	U_{iiii} = g\int\mathrm{d}\vecr\, f^4_i(\vecr) ,
\end{equation}
which under the substitutions of Eqs.~\eqref{cond_n} and \eqref{cond_w} becomes
\begin{align}
	\label{ostwopar}
	U_{iiii}^{W}&=g\int\mathrm{d}\vecr~ |W_i(\vecr,\vecr)|^2 ,
\\
	\label{oswannier}
	U_{iiii}^{w}&=g\int\mathrm{d}\vecr ~|w_i(\vecr)|^{4} .
\end{align}

The second group corresponds to terms with operators $\hat{a}_i^\dagger \hat{a}_i^\dagger \hat{a}_j \hat{a}_j, (i\neq  j)$, which describe the joint tunneling of two particles between two neighbouring lattice sites, i.e.~the particles hop together from one lattice site to another. The coupling amplitudes associated with this process are given by
\begin{equation}
	\label{ptU}
	U_{iijj} = g\int\mathrm{d}\vecr\, f^2_i(\vecr)f^2_j(\vecr) ,
\end{equation}
and become after substitution
\begin{align}
	\label{pttwopar}
	U_{iijj}^{W}&=g\int\mathrm{d}\vecr~ |W_i(\vecr,\vecr)|  |W_j(\vecr,\vecr)| ,
\\
	\label{ptwannier}
	U_{iijj}^{w}&=g\int\mathrm{d}\vecr ~|w_i(\vecr)|^{2} |w_j(\vecr)|^{2} .
\end{align}

The next effect is associated with terms including $\hat{a}_i^\dagger \hat{a}_j^\dagger \hat{a}_i \hat{a}_j$, and it can be interpreted as two indistinguishable processes:
the interaction between particles at neighbouring sites or cross tunneling of particles.
As these processes only involve a single particle at each site, one gets $U^{W}_{ijij} = U^{w}_{ijij} = U^{w}_{iijj}$.

Finally, the last effect is associated with terms including $\hat{a}_i^\dagger \hat{a}_i^\dagger \hat{a}_i \hat{a}_j$, which
describes single-particle tunneling between an empty and an already occupied neighbouring trap. The coupling amplitudes for this process are given by
\begin{equation}
	\label{mixU}
	U_{iiij} = g\int\mathrm{d}\vecr\, f^3_i(\vecr)f_j(\vecr) ,
\end{equation}
which, after the substitutions, become
\begin{align}
	\label{mixtwopar}
	U_{iiij}^{W}&=g\int\mathrm{d}\vecr~ |W_i(\vecr,\vecr)| |w_i(\vecr)| |w_j(\vecr)| ,
\\
	\label{mixwannier}
	U_{iiij}^{w}&=g\int\mathrm{d}\vecr ~|w_i(\vecr)|^{3} |w_j(\vecr)| .
\end{align}

\section{Results and discussions}
\label{sec:results}

\begin{figure*}[tb]
\begin{minipage}{0.49\textwidth}
\centering (a) $V_0=10E_r$ \\ \includegraphics[width=0.95\textwidth]{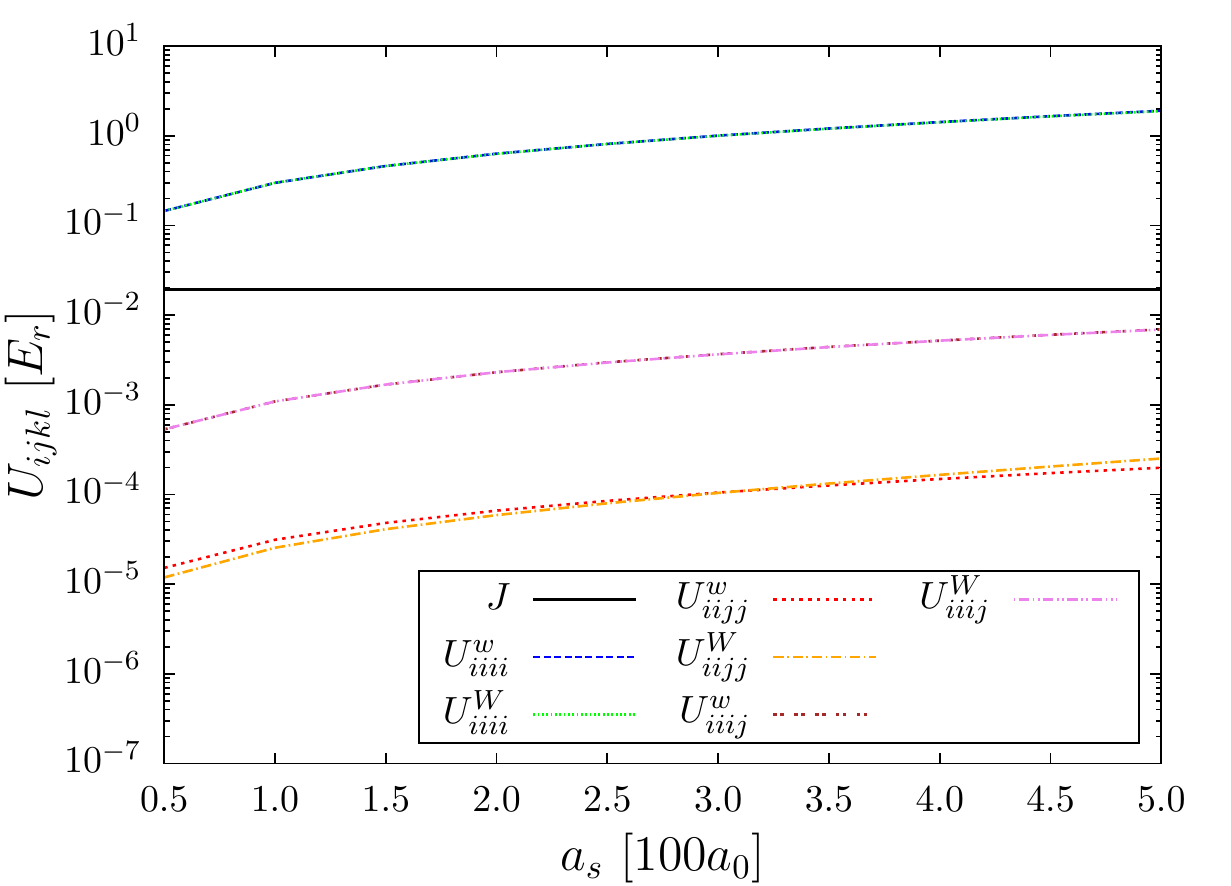}
\end{minipage}
\begin{minipage}{0.49\textwidth}
\centering (b) $V_0=20E_r$ \\ \includegraphics[width=0.95\textwidth]{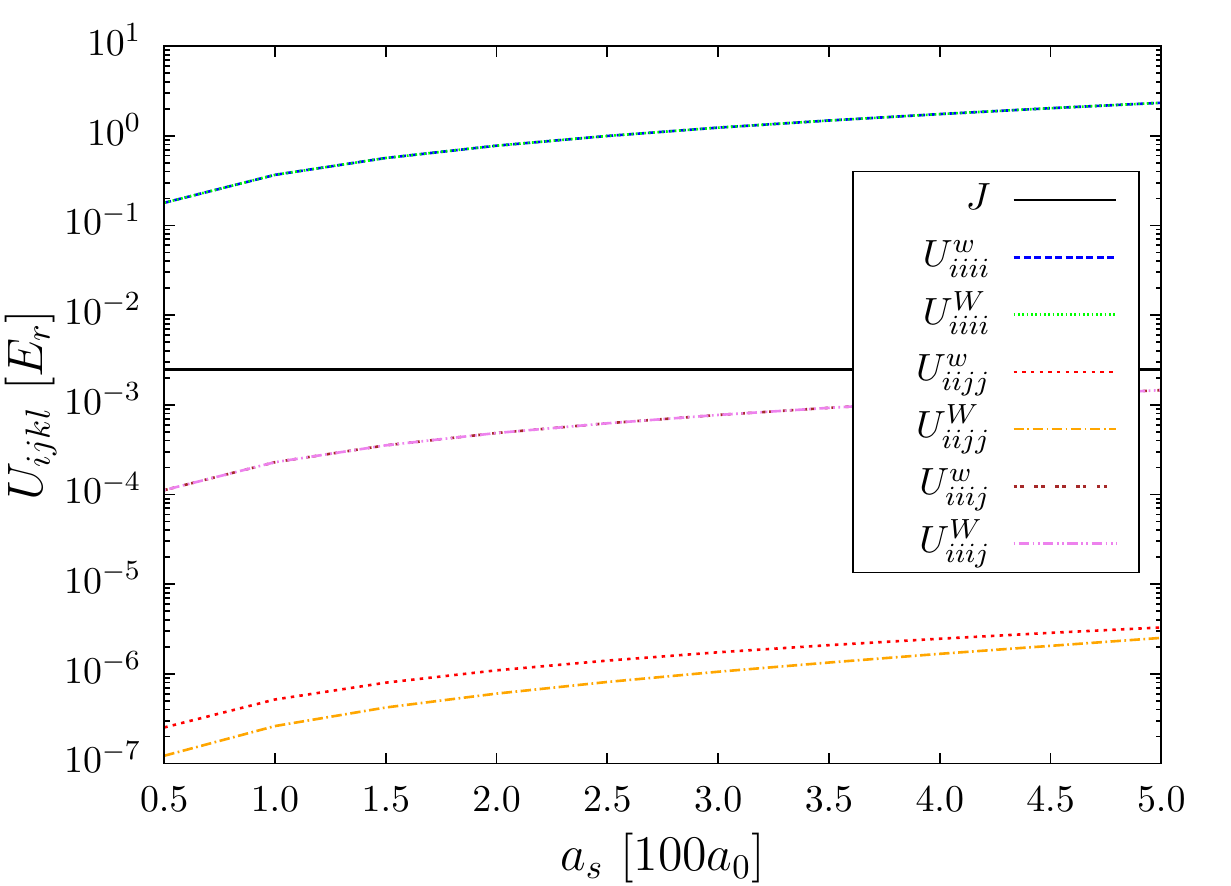}
\end{minipage}
\caption{\label{fig:s20+s10}
Dependence of the Bose--Hubbard parameters (plotted logarithmically)
on scattering length $a_s$ (in units of $100 a_0$)
for lattice depths (a) $V_0=10E_r$  and (b) $V_0=20E_r$.
The curves correspond to
$U_{iiii}^{w}$ (dashed blue),
$U_{iiii}^{W}$ (dotted green),
$J$ (solid black),
$U_{iiij}^{w}$ (double-dotted dark red),
$U_{iiij}^{W}$ (dot-dash-dotted pink),
$U_{iijj}^{w}$ (small-dashed red), and
$U_{iijj}^{W}$ (dashed-dotted orange). For the numerical calculation nine traps have been taken into account.}
\end{figure*}

In the following we will numerically compute and compare the interaction parameters for the single-particle and the two-particle Wannier function approach. To avoid complications from the regularized delta function in three dimensions, all calculations are done in one dimension, assuming a tight harmonic confinement of the atoms in the transverse direction (of frequency $\omega_\perp$). However, all
calculations are conceptually straightforward to extend to higher dimensions. Adjusting the coupling constant $g$ to one dimension can be done via
$g_\text{1D}=-\frac{2\hbar^2}{ma_\text{1D}}$,
with $a_\text{1D}=-\frac{d_\perp^2}{2a_s}\left(1-C\frac{a_s}{d_\perp}\right)$,
where $C \simeq 1.4603$ and $d_\perp=\sqrt{\frac{2\hbar}{m\omega_\perp}}$ ~\cite{olshanii98}. In the following we choose $\omega_\perp=2\pi\times 10^4$ Hz.

The results for  two different values of the scattering length ($a_s=100\,a_0$ and $a_s=400\,a_0$) and as a function of the lattice depth are shown in Fig.~\ref{fig:depths}. It can be seen that the overlap integrals $U_{iiii}$, which describe the on-site interaction, are generally in good agreement with each other for both approaches. The biggest deviations appear for shallow lattices (see Fig.~\ref{fig:depths}(c)), where 
$U_{iiii}^{W}$ is smaller than $U_{iiii}^{w}$. The difference stems from the fact that the repulsive interaction leads to a broadening of two particle density and consequently a reduction in its maximal amplitude, which directly translates into a smaller magnitude of the interaction coefficient for the two-particle Wannier approach. For deeper lattices, i.e.~larger potential energies, the broadening is reduced and the two quantities have similar values. The crossing between $U_{iiii}$ and $J$, which is visible in the inset of Fig.~\ref{fig:depths}(a), corresponds to the parameter range where tunneling starts to dominate over the interaction effects. Since at the crossing point the two relevant values of $U_{iiii}$ differ by about $10\%$, an effect on the Mott-transition point can be expected.

Similar differences between the two methods can also be noted for the overlap integrals for the correlated pair tunneling, $U_{iijj}$, where for shallow lattices the integral based on the two-particle Wannier functions is larger than the one based on the single-particle functions.
Here the extended size of the localised functions due to the repulsive interactions leads directly to a larger overlap between neighboring sites. On the other hand, for deeper lattices, the pair-tunneling coupling calculated from the two-particle functions becomes an order of magnitude smaller than that from the single-particle functions. This is due to the fact that even at higher lattice depths the single particle Wannier function density and the two particle density have different behaviour in their tails, although their bulk density becomes almost identical.
In this regime, the magnitude of the tail of the single particle Wannier density is higher than the one of the two particle density, leading to a larger overlap between neighboring densities, and thus to higher values of $U_{iijj}$ (see also  Fig.~\ref{fig:depths}(d)).
Finally, the density dependent couplings $U_{iiij}$ show a difference for shallow lattices, which can be explained in the same way as for the interaction terms above (see Fig.~\ref{fig:depths}(e)).

These results are consistent with the situation where the interaction strength is changed while keeping the lattice depth constant (see Fig.~\ref{fig:s20+s10}).
The on-site interaction and interaction-mediated tunneling terms, $U_{iiii}$ and $U_{iiij}$, do not show much difference between the two methods, but the two-particle tunneling coupling, $U_{iijj}$ is much more severely affected. 
For a comparatively deep lattice ($V_0=20E_r$, Fig.~\ref{fig:s20+s10}(b)) the two-particle tunneling amplitude calculated using the two-particle Wannier approach increases faster than the one based on the single-particle Wannier functions, 
and the two methods do not coincide anywhere in the plotted parameter regime. However, for a shallower lattice ($V_0=10E_r$, Fig.~\ref{fig:s20+s10}(a)) a crossing can be seen, as the two curves associated to $U_{iijj}$ are closer together. This leads to the conclusion that the effects of the interactions can have significant influence on the parameters of the Bose--Hubbard model, and should be taken into account in particular in metrology experiments. It also provides justification for the use of extended Bose--Hubbard models~\cite{rashi1,rashi2}, which take the two-particle tunnelling and the cross tunnelling terms into account~\cite{ebhm:lewenstein,benseny,reshodko}.

\section{Possible applications and conclusions}
\label{sec:conc}

To summarize, we have calculated the parameters for the Bose--Hubbard model by consistently including on-site density effects. This was done by replacing the commonly used single-particle Wannier functions by two particle Wannier functions, which result in a broadening of the density due to repulsive interactions. Given the experimental control parameter of the optical lattice depth and the scattering lengths, we have shown that in certain regimes the Bose--Hubbard parameters show substantial deviation from the results using single-particle Wannier functions and that terms such as the correlated pair tunnelling can be become important, even though they are usually neglected.

These results are hence of principle interest for current and future experiments in the field of ultracold atoms in optical lattices, especially to account for non-uniform shifts in atomic clock frequencies due to the collision of atoms.  In a recent experiment by Campbell {\it et al.~}\cite{Campbell:06}, the atomic clock shift of $^{87}$Rb was measured, and found to decrease with increasing number of atoms per site. Other works have also shown that the clock frequency shift is directly proportional to the onsite interaction strength~\cite{Cornell:02,Baym:96}. When calculated using single particle Wannier functions, the onsite interaction term is independent of the occupancy of lattice sites, and hence cannot explain the decrease of the clock shift with increasing occupancy. However, the presented technique takes into account the effect of repulsive interactions implicitly, and the resulting broadening of the two-particle single-site density and the decrease of the magnitude of onsite interaction term $U_{iiii}$, can explain the decrease of clock shift.

\end{document}